# Performance of Kriging Based Soft Classification on WiFS/IRS- 1D image using Ground Hyperspectral Signatures

Sumanta Kumar Das and Randhir Singh

*Abstract*— Hard/soft classification techniques are the conventional ways of image classification on satellite data. These classifiers have number of drawbacks. Firstly, these approaches are inappropriate for mixed pixels. Secondly, these approaches do not consider spatial variability. Kriging based soft classifier (KBSC) is a non-parametric geostatistical method. It exploits the spatial variability of the classes within the image. This letter compares the performance of KBSC with other conventional hard/soft classification techniques. The satellite data used in this study is the Wide Field Sensor (WiFS) from the Indian Remote Sensing Satellite -1D (IRS-1D). The ground hyperspectral signatures acquired from the agricultural fields by a hand held spectroradiometer are used to detect subpixel targets from the satellite images. Two measures of closeness have been used for accuracy assessment of the KBSC to that of the conventional classifications. The results prove that the KBSC is statistically more accurate than the other conventional techniques.

## I. INTRODUCTION

CONVENTIONAL ways of image classification of satellite data are based on class discrimination using hard/soft classification techniques. Hard classification technique has a number of drawbacks that limit its practical applications. It works at a pixel level without allowing estimation at subpixel level. Most of the hard classifiers rely on a Gaussian distribution for the spectral signatures of the training data that often exhibits a non Gaussian distribution. Hard classifiers do not quantify the likeliness (or probability) that a pixel actually belongs to a pre-defined class; neither do they consider spatial variability.

An alternative to the hard classification is soft classification that has been widely used for actual assessment of class proportions of a mixed pixel. Most commonly used soft classification techniques are based on pixel unmixing (e.g. Linear mixing model (LMM) [1]), soften version of maximum likelihood [2], fuzzy logic [3-4], neural network [5], kernel nonparametric method [6], varying-time-regression model [7], multilogit model [7]. The LMM is the basic and widely used pixel-unmixing technique to decompose mixed pixels into a collection of distinct end members along with their abundances. This model has some limitations: 1) "condition of identifiability"[8]; 2) assumption for known and constant end members spectra [8]; 3) assumption of linear mixing [9]; 4) high correlation of hyperspectral bands [10-11]. Some of the feature extraction (or dimensionality reduction) techniques that have been used to improve the LMM are principal components transform, discrete wavelet transform [11-12], Fisher linear discriminat transform, spectral band selection and singular value decomposition (SVD)[13], use of vicinal information [8]. Incidentally, none of these classifiers consider the spatial variability; even-though this is an important factor of satellite images.

Satellite sensors collect data at a range of resolutions. These are spatially auto correlated. A classifier that uses spatial information ([14-15]) of the satellite data is always more favorable than the conventional classifiers. In addition, if the classifier can predict the abundances of end members along with their spatial location within the pixel then it becomes much superior. Moreover, if the classifier is scale free, i.e. it works on both small (downscale or subpixel) as well as large (upscale or group of pixels) scales then it seems to be most useful classifier. To a large extent the kriging based soft classifier (KBSC) satisfies all of these criteria [15-16].

Every feature on the earth has its own unique spectral signatures in the electromagnetic spectrum. The ground hyperspectral signatures (GHS) are often collected and stored in a spectral library so that they can be used for comparing the hyperspectral signatures from satellite imagery for pixel identification. Generally the GHS are collected with a high precision in a laboratory like environment, where as the satellite data are not free from atmospheric scattering and absorption. As a result before applying the GHS for pixel identification, the satellite data is being corrected for the atmospheric scattering and absorption.

The present letter evaluates the performance of the KBSC, in comparison to other conventional classifiers such as maximum likelihood, Bayesian, Dempster-Shafer, fuzzy classifiers. So far the KBSC has been used for mineral mapping from the hyperspectral data like the Airborne Visible/Infrared Imaging Spectrometer (AVIRIS) [16] or for the simulated image data [17] successfully; still its performance for extracting the class features (specially agricultural) from nonhyperspectral coarse spatial resolution data like the WiFS / IRS 1-D has not been explored previously.

The main objective of this study is to evaluate the performance of KBSC using the GHS in order to perform subpixel target detection. The Performance of the KBSC is validated by applying this classifier for area estimation of agricultural crops from a nonhyperspectral sensor data like WiFS from IRS 1D satellite.

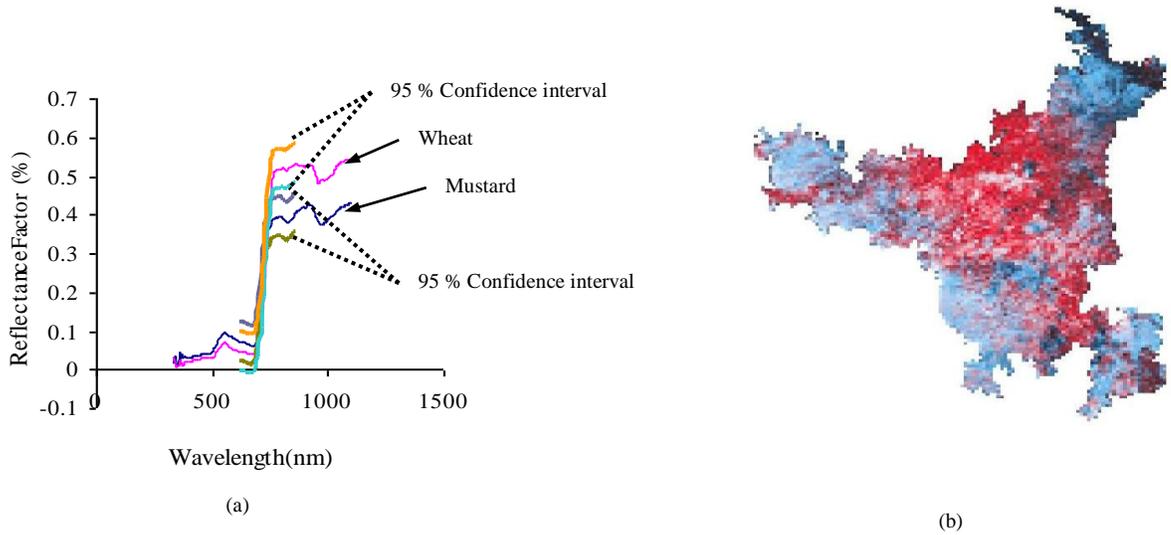

Fig.1. (a) Spectral signature curve of wheat and mustard collected from field experiments using a handheld spectroradiometer , (b) False Color Composite of WiFS of 16[th] Feb.1998 using red and infrared bands.

## II. MATERIALS AND METHODS

The study area is the state of Haryana in India, which is located between 74° 25′ to 77° 38′ E longitudes and 27° 40′ to 30° 55′ N latitudes. The WiFS data from the IRS-1D [18] is used in this study; it has spatial resolution of 188 m that covers the entire state (Fig. 1(b)). The date of acquiring the data is the February 16, 1998. The WIFS has two spectral bands; one in visible red band (RED) i.e.620-680 nm and the other in infrared region (IR) i.e.770-860 nm and its swath is 810 km. Since the pixel size of data is large (around 188 m × 188 m), many pixels are of mixed composition. The satellite data is georeferenced using ground reference points obtained from a global positioning system (GPS).

Main agricultural crops of the state during this season are winter wheat (*Triticum aestivum L.*) and mustard (*Brassica juncea L.*). After visually observing the satellite data, it is found that water, buildings and road features are easily identifiable from the satellite data. However, it is not easy to distinguish wheat and mustard pixels from the pixels covered by vegetations. Though, most of the satellite's pixels are of mixed type, a soft classification of the satellite data will be appropriate. Laboratory based spectral signatures curves are developed and used to train the satellite data. Spectral signatures of wheat and mustard (from 75 wheat and 65 mustard fields) are acquired from agricultural fields spread across the study area. The hyperspectral data are collected using an Analytical Spectral Device Fieldspec® handheld spectroradiometer [19], in order to obtain pure end member signatures, which have 700 spectral bands that are sampled at 1 nm over the range of 400 to 1100 nm with a spectral resolution of 10 nm. A 25° instantaneous-field-of-view foreoptic is used, the instrument is set to average ten signatures to produce each sample signature, and the sensor is held nadir at approximately 4 ft above the vegetation canopy. Signatures are collected for 2 classes, namely: 1) winter wheat (*Triticum aestivum L.*); and 2) mustard (*Brassica juncea L.*).

The National Sample Survey Organization (NSSO), India, collects the crop cutting experiment (cce) and GHS data under the general crop estimation surveys (GCES)[20]. A stratified multistage random sampling design is adopted in these surveys where the blocks (two or three districts together) constitute the strata. A sample of villages is selected from different strata in proportion to the area under crop, based on the past year data. From each selected village, two fields are selected randomly and from each field, a plot of fixed size, generally measuring 10 m × 5 m; is selected. Geographical locations of each sampled fields are recorded by a GPS.

All the spectra collected from sampled fields of wheat and mustard, are averaged to constitute a single representative spectrum (Fig. 1(a)). The sampled fields are not necessarily of pure classes as few (approximately 40 %) of them are mixed. The relative proportions of different ground classes in mixed fields are recorded by the observer's eye estimations.

The WiFS data taken from IRS −1D is not free from the atmospheric effects. The dark object subtraction model (DOS) developed by Chavez 1998 [23] is applied for atmospheric correction of the satellite data. This algorithm is available with the ATMOSC module of commercial image processing software IDRISI [21]. In order to correlate the hyperspectral reflectance field data with the satellite's DNs (Digital Number), the hyperspectral reflectance data are reduced to satellite's DN using two parameters (gain and bias provided with the satellite data) using the following equations [22]:

$$L(i) = [R(i) * (E_{sun} \cdot \sin(SE))]/(\pi \cdot d^2) \quad (1)$$
$$DN(i) = [L(i) - bias(i)]/gain(i) \quad (2)$$

where $L$: at satellite radiance in mW cm$^{-2}$sr$^{-1}$μm$^{-1}$; $i$ = band number; $R$: at-satellite reflectance (unit less); $E_{sun}$ = mean solar exoatmospheric irradiance in mW cm$^{-2}$sr$^{-1}$μm$^{-1}$; $SE$: sun elevation angle (in degrees) and d: earth-sun distance in astronomical unit. Gain in mW cm$^{-2}$sr$^{-1}$μm$^{-1}$ and bias in mW cm$^{-2}$sr$^{-1}$μm$^{-1}$, values are provided with the header file of the satellite data.

Let the digital numbers (DNs) of a remotely sensed satellite imagery are the realizations of independent and identically distributed (*iid*) spatial random variables (rv) (Z (**x**)) of a random process; where **x** is the location. Two stationary

**STEP I:** Suppose a block of size 3x3 (or more) is selected from the band A (620-680 nm) and band B (770-860 nm) data of WiFS from the IRS-1D. Number in the cell indicates the digital numbers (DNs).

BAND A
| 60 | 59 | 60 |
|---|---|---|
| 63 | 53 | 36 |
| 58 | 45 | 40 |

BAND B
| 55 | 32 | 30 |
|---|---|---|
| 57 | 70 | 125 |
| 52 | 115 | 85 |

**STEP II:** Extract the upper $(U_i)$ and lower $(L_i)$ limits of the DN in the spectral band $i$ (=A, B) from the ground hyperspectral signatures. e.g. CLASS 1: WHEAT  $L_A < DN_A < U_A$; $L_B < DN_B < U_B$

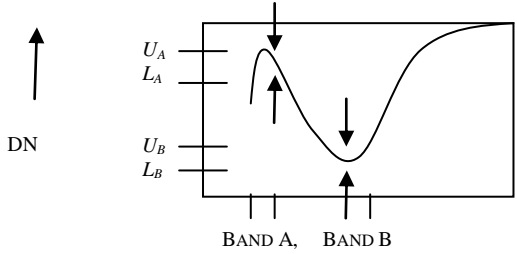

**STEP III:** Transform band data to binary data (1 if below the threshold and 0 if above the threshold)

BAND A
| 1 | 1 | 0 |   | 0 | 0 | 1 |
|---|---|---|---|---|---|---|
| 1 | 1 | 0 |   | 0 | 1 | 1 |
| 0 | 0 | 0 |   | 1 | 1 | 1 |
UPPER   LOWER

BAND B
| 1 | 1 | 0 |   | 0 | 0 | 1 |
|---|---|---|---|---|---|---|
| 1 | 0 | 0 |   | 1 | 0 | 1 |
| 1 | 0 | 0 |   | 0 | 1 | 1 |
UPPER   LOWER

**STEP IV:** Apply kriging through variogram model fitting; (use any output block size)

BAND A
| 1 | 0.9 | 0.6 | 0.1 |   | 0 | 0 | 0.6 | 0.9 |
|---|---|---|---|---|---|---|---|---|
| 1 | 0.9 | 0.5 | 0.1 |   | 0 | 0.2 | 0.8 | 1 |
| 0.8 | 0.7 | 0.2 | 0 |   | 0 | 0.2 | 0.8 | 1 |
| 0.4 | 0.3 | 0 | 0 |   | 0.8 | 0.9 | 1 | 1 |
UPPER      LOWER

BAND B
| 1 | 0.9 | 0.4 | 0.1 |   | 0 | 0 | 0.4 | 0.8 |
|---|---|---|---|---|---|---|---|---|
| 0.9 | 0.7 | 0.5 | 0.1 |   | 0 | 0.1 | 0.3 | 0.8 |
| 1 | 0.5 | 0.1 | 0 |   | 0.8 | 0.5 | 0.6 | 1 |
| 0.9 | 0.6 | 0.2 | 0 |   | 0.4 | 0.9 | 0.9 | 1 |
UPPER      LOWER

**STEP V:** Integrate upper and lower limits (degree of fuzziness of DN values within spectral range; from the equation 5)

BAND A
| 1 | 0.9 | 0.84 | 0.91 |
|---|---|---|---|
| 1 | 0.92 | 0.9 | 1 |
| 0.8 | 0.76 | 0.84 | 1 |
| 0.4 | 0.93 | 1 | 1 |

BAND B
| 1 | 0.9 | 0.64 | 0.82 |
|---|---|---|---|
| 0.9 | 0.73 | 0.65 | 0.82 |
| 1 | 0.75 | 0.64 | 1 |
| 0.94 | 0.96 | 0.92 | 1 |

**STEP VI:** Calculate the joint probability (from the equation 6)

| 1 | 0.81 | 0.5376 | 0.7462 |
|---|---|---|---|
| 0.9 | 0.6716 | 0.585 | 0.82 |
| 0.8 | 0.57 | 0.5376 | 1 |
| 0.376 | 0.8928 | 0.92 | 1 |

Fig. 2. Pictorial description of the KBSC

$$\hat{Z}(x_0) = \sum_{l=1}^{k} \lambda_l Z(x_l) \qquad (3)$$

where

$$\sum_{l=1}^{k} \lambda_l = 1 \qquad (4)$$

The entire methodology of the KBSC is described through a schematic diagram in Fig. 2. In most of the hyperspectral imagery classification, basic step is to identify the key bands where a particular class feature is highly distinguishable. This step is not required here because vegetation feature is highly distinguishable in the RED and IR bands of the WiFS/IRS 1-D.

Let $U_i$ and $L_i$ refer to the upper and lower confidence intervals of DN, at $\alpha$ % confidence level for band $i$. These are defined as: $\left( \bar{x}_i \pm t_{\alpha,n} \frac{s_i}{\sqrt{n}} \right)$, where $\bar{x}_i$ and $s_i$ are the mean and standard deviation of DNs for band $i$ obtained from the ground data, $t_{\alpha,n}$ is the area under the Student's $t$- distribution with $\alpha$ level of significance and $n$ is the degrees of freedom (number of bands of the GHS within the band limits of WiFS). By interpolating the binary maps for the upper limit, we obtain a map representing the probability that the value of a training block lies below the indicated upper limit (say event $E_1$). The interpolation for the lower limit results in a map representing the probability that the block value lies above the lower limit (say event $E_2$). Combining these we obtain the probability that the block DN value is higher than the lower limit and smaller than the upper limit (e.g., the probability of a block having a spectral DN value in between the predefined range corresponding to the spectral response of the crop of interests). Though each DN is assumed as a realization of *iid* rv $Z(\mathbf{x})$, $E_1$ and $E_2$, are independent, the compound probability event is simply defined as

$$Pr\{map_i\} = Pr\{E_1 \cup E_2\} = Pr\{E_1\} + Pr\{E_2\} - Pr\{E_1\} Pr\{E_2\} \qquad (5)$$

The KBSC has the flexibility to define the size of a pixel to produce output and it is not necessary to be of the same size as the input pixel size. This provides a means of extrapolating to areas larger than the pixel size or interpolating to areas smaller than the pixel size. Repeating this procedure for all key bands results in a set of probability maps (one for each band), which can be integrated by calculating the joint probability that is a measure for the likeliness that a pixel belongs to a certain class. The joint probability is obtained as (assuming that each bands are linearly independent)

$$Pr\{joint\} = Pr\{map_i\}Pr\{map_j\}Pr\{map_k\}...Pr\{map_b\} \qquad (6)$$

where $map_i$ to $map_b$ are the maps representing probability maps for key bands $i$ to $b$ used. This image is used as input for the classification. By setting tolerances on the minimum probability acquired for each class, pixel can be classified with a predefined accuracy. The level of accuracy of the classification is proportional to the average proportion of a block area chosen as a threshold level.

The performance of the KBSC is compared with four conventional hard/soft classifiers. These algorithms are available with the MAXLIKE, BAYCLASS, BELCLASS and assumptions are made to allow statistical inference: 1) expected value ($m=E\{Z(\mathbf{x})\}$) exists and is independent of $\mathbf{x}$; 2) The inter-dependence between any two-point locations (e.g. variogram; $2\gamma(\mathbf{h}) = E[\{Z(\mathbf{x}) - E(\mathbf{x}+\mathbf{h})\}^2]$) is expressed as a function of lag ($\mathbf{h}$; separating distance and direction). The best linear unbiased estimator (BLUE) of the value of the variable at any unknown location ($\mathbf{x_0}$) using the values of known locations $x_1,...x_k$, can be obtained by kriging estimator

FUZZYCLASS modules of commercial image-processing software IDRISI 32 ([21]). The MAXLIKE is the maximum likelihood classifier. It assigns each pixel to the most likely class. The BAYCLASS employs Bayesian probability theory to express the degree of membership of a pixel to any class. The BELCLASS employs Dempster-Shafer theory of evidence using belief functions and plausibility reasoning. It is used to combine separate pieces of information (evidence) to calculate the probability of an event. It is a variant of the BAYCLASS. It estimates the belief interval, as a measure of classification uncertainty. The belief interval is the difference between belief (the degree to which evidence provides support for a hypothesis) and plausibility (the degree to which the evidence does not refute that hypothesis). The FUZZYCLASS is based on supervised fuzzy classifier developed by Wang in 1990[4].

The FUZSIG module is used to develop the fuzzy signatures (the fuzzy mean and the fuzzy covariance matrix [4]) from training pixels. The FUZSIG creates the signatures from information contained in the remotely sensed images from the training samples. These signatures are used to perform a supervised classification of the remotely sensed imagery using the BAYCLASS, BELCLASS and the FUZZYCLASS modules. For the KBSC, class proportions of each subpixel locations are determined by kriging. The KRIGECLASS (algorithm developed for the KBSC) uses kriging based class probability (from equation 6) and the grid distance ($h$) for the variogram model fitting to develop KBSC results.

As the large portions of the image are composed of mixed pixel, an accuracy assessment based on pure pixel will not provide full or adequate description of classifications performance.

The Linear Imaging Self Scanner (LISS III, [18]) data is used for accuracy assessment, which is obtained from the same satellite for the same day and location. The LISS III data has a spatial resolution of 23.5 m. The LISS III data is classified using MAXLIKE, BAYCLASS, BELCLASS, FUZZYCLASS and KRIGECLASS. The soft classified LISS III data is up scaled (passing through a mean filter of $8 \times 8$ window) to 188 m resolution to compare with the soft classified WiFS data.

Foody *et al.* [3] have shown that usual method of accuracy assessment (confusion matrix) is not capable of measuring the accuracy of soft classification. The closeness of probability distribution of different classifiers on the WiFS data and the LISS III data is done by three criterions: means square error, cross entropy and correlation analysis.

*A. Mean Square Error:*

$$S = \frac{1}{C} \sum_{c=1}^{C} (e_{1c} - e_{2c})^2 \qquad (7)$$

where $e_{1c}$ is the proportion of class $c$ in a pixel from the LISS III data and $e_{2c}$ is the proportion of the class $c$ in a pixel from the WIFS data, $c$ denotes the number of classes in the data set.

*B. Cross entropy:*

$$D(f_1, f_2) = -\sum_x f_1(x) \log_2 f_2(x) + \sum_x f_1(x) \log_2 f_1(x) \qquad (8)$$

$f_1$ is the proportion from the LISS III data and $f_2$ is the proportions from the WiFS data.

## III. RESULTS AND DISCUSSION

Table I shows the distribution of $S$ and $D$ for different classifiers. A low value of $S$ and $D$ is an indicative of good representation of classified data to that of the true data. Examining the Table I it is found that the KBSC is a better classifier compared to other conventional classifiers. The WiFS data is classified at subpixel level by varying the grid distances ($h$ = 23.5, 47 m, 188 m, Table I). The maximum efficiency of the KBSC is obtained when the grid distance ($h$=188 m) is equal to the spatial resolution of the image. The efficiencies of the KBSC go down (still comparable to the other methods) when classification is made above/below the spatial resolution of the satellite data.

Table II shows the values of the correlation coefficients ($R^2$) between the proportions of classes on the WiFS data and proportions of classes on the LISS III data. Analyzing these values of $R^2$ for different soft classifiers, it can be said that the accuracy of the KBSC is significantly better from other classifiers.

A comparison of the percentage deviation (PD) of MAXLIKE, BAYCLASS, BELCLASS and KRIGECLASS area estimates with the usual GCES estimate is given in Table III. According to the GCES based acreage estimates, wheat and mustard cover 45.72% and 13.99% of the total geographical area (437, 3861 ha) of the state. The results show that the entire satellite based area estimation techniques underestimate the area as compared to GCES estimates. For wheat in all the cases, it is less than 16 % and in case of mustard, it is less than 18 %. The overall PDs of MAXLIKE were over 15.74% to 17.41% for wheat and mustard respectively. If the PDs of KRIGECLASS, BAYCLASS and BELCLASS estimates are compared it is found that KBSC estimates have less PD. The PD of KRIGECLASS based estimate with varying grid distances ($h$) is also studied.

## IV. CONCLUSION

The results verify that the KBSC performs well compared to other conventional classifiers for classifying nonhyperspectral satellite data like WiFS from IRS-1D with the help of GHS. In addition, KBSC is scale independent that means this can be used for both within (subpixel) or beyond (macropixel) the spatial resolution of the satellite data. Although, KBSC performs better for subpixel level in compare to the macropixel level.

TABLE I
THE MEAN, MEDIAN AND STANDARD DEVIATION (SD) OF THE DISTRIBUTIONS OF TWO CLOSENESS MEASURES (S & D) FOR DIFFERENT CLASSIFICATIONS.

| Classifier | Measure of closeness | | | | | |
|---|---|---|---|---|---|---|
| | $S$ (Mean Square Error) | | | $D$ (Cross Entropy) | | |
| | Mean | Median | SD | Mean | Median | SD |
| MAXLIKE | 0.0964 | 0.0946 | 0.0665 | 0.0845 | 0.0845 | 0.0645 |
| BAYCLASS | 0.0845 | 0.0856 | 0.0556 | 0.0712 | 0.0710 | 0.0654 |
| BELCLASS | 0.0766 | 0.0761 | 0.0751 | 0.0651 | 0.0695 | 0.0514 |
| FUZZYCLASS | 0.0689 | 0.0678 | 0.0535 | 0.0414 | 0.0542 | 0.0602 |
| KRIGECLASS | | | | | | |
| $h$ = 23.5 m | 0.02053 | 0.02075 | 0.0262 | 0.0210 | 0.0211 | 0.0247 |
| $h$ = 1000 m | 0.03063 | 0.03119 | 0.0268 | 0.0346 | 0.0334 | 0.0390 |
| $h$ = 188 m | 0.01766 | 0.01824 | 0.0283 | 0.0112 | 0.0102 | 0.0259 |

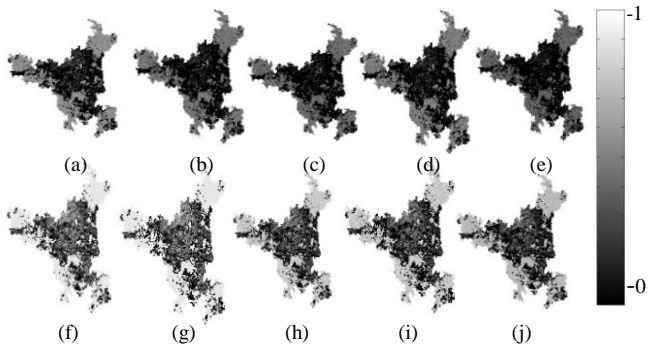

Fig. 3. Images of two closeness measures: (a-e) the mean square error ($S$, first row) & (f-j) the cross entropy ($D$, second row) using MAXLIKE, BAYCLASS, BELCLASS, FUZZYCLASS, KRIGECLASS respectively.

TABLE II
CORRELATION COEFFICIENTS (STATISTICALLY DIFFERENT AT $\alpha$ = 0.05) AS A MEASURE OF ASSOCIATION BETWEEN WIFS CLASSIFIED DATA AND LISS III CLASSIFIED DATA

| CLASSIFIER | MXL | BAY CLASS | BEL CLASS | FUZZY CLASS | KRIGE CLASS |
|---|---|---|---|---|---|
| Wheat | 0.75 | 0.78 | 0.76 | 0.76 | 0.82 |
| Mustard | 0.76 | 0.75 | 0.76 | 0.73 | 0.82 |
| Other Vegetation | 0.76 | 0.76 | 0.72 | 0.75 | 0.71 |
| Shallow Water | 0.75 | 0.72 | 0.75 | 0.76 | 0.81 |
| Deep Water | 0.76 | 0.72 | 0.72 | 0.76 | 0.81 |
| Buildings | 0.75 | 0.82 | 0.72 | 0.72 | 0.81 |
| Road | 0.75 | 0.75 | 0.78 | 0.75 | 0.81 |

TABLE III
ESTIMATED CROP AREA UNDER WHEAT AND MUSTARD FROM WIFS/ IRS 1D FOR THE STATE OF HARYANA, INDIA DURING FEBRUARY 16, 1998 (ha)

| Classifier | Wheat | % Dev[1] | Mustard | % Dev[2] |
|---|---|---|---|---|
| GCES | 2000,000 | | 612,000 | |
| MAXLIKE | 1,684,582 | -15.7709 | 508,235 | -16.9551 |
| BAYCLASS | 1,844,554 | -7.7723 | 522,576 | -14.6118 |
| BELCLASS | 1,845,392 | -7.7304 | 528,348 | -13.6686 |
| KRIGECLASS | | | | |
| $h$ = 1000 m | 1,858,456 | -7.0772 | 536,035 | -12.4126 |
| $h$ = 235 m | 1,894,683 | -5.2658 | 557,525 | -8.9011 |
| $h$ = 188 m | 1,898546 | -5.0727 | 585,341 | -4.3560 |

[1] Calculated as: % Deviation= [GCES+/Wheat)×100]-100;
[2] Calculated as: % Deviation= [GCES+/Mustard)×100]-100;